%% file: main_sigmetrics.tex
\documentclass[sigconf]{acmart}
\input{headers}
\newcommand{\acc}{\texttt{Accuracy}\xspace}
\newcommand{\sobol}{\texttt{Sobol}\xspace}
\newcommand{\adjacent}{\texttt{Adjacent}\xspace}

\newcommand{\extrema}{\texttt{Extrema}\xspace}
\newcommand{\shm}{\texttt{Shmembench}\xspace}

\newcommand{\add}{\texttt{Add}\xspace}

\newcommand{\triad}{\texttt{Triad}\xspace}

\newacronym{rpm}{RPM}{Requests per Minute}
\newcommand{\gpt}{\texttt{GPT-5}\xspace}
\newcommand{\gemini}{\texttt{gemini-2.5-pro}\xspace}
\newcommand{\llama}{\texttt{Llama3.1:70B}\xspace}

\newcommand{\sysname}{\textsc{Optimas}\xspace}

\AtBeginDocument{%
  }

\setcopyright{none}
\acmConference[arXiv Preprint]{arXiv Preprint}{2026}{}

\begin{document}

\title{\sysname: An Intelligent Analytics-Informed Generative AI Framework for Performance Optimization}




\author{Mohammad Zaeed}
\affiliation{%
  \institution{Texas State University}
  \city{San Marcos}
  \state{Texas}
  \country{USA}}
\email{cup7@txstate.edu}

\author{Tanzima Z. Islam}
\affiliation{%
  \institution{Texas State University}
  \city{San Marcos}
  \state{Texas}
  \country{USA}}
\email{tanzima@txstate.edu}

\author{Vladimir Indic}
\affiliation{%
  \institution{University of Novi Sad}
  \city{Novi Sad}
  \country{Serbia}}
\email{vladaindjic@uns.ac.rs}

\renewcommand{\shortauthors}{Anonymous et al.}

\begin{abstract}
\input{short-abstract}
\end{abstract}



\keywords{HPC, Performance profiling and bottleneck detection, Performance Optimization, Large Language Models, Multi-Agent System, CUDA}


\maketitle
\sloppy

\section{Introduction}
\label{sec:intro}
\input{intro-kdd}

\section{Our Approach}
\label{sec:approach}
\input{approach3}

\section{Implementation Details and Usage}
\label{sec:framework}
\input{framework5-kdd}

\section{Evaluation Setup}
\label{sec:setup}
\input{setup2}

\section{Results}
\label{sec:results}
\input{results3}
\section{Related Work}
\label{sec:related-work}
\input{related-work}

\section{Conclusions}
\label{sec:conclusions}
\input{conclusions}

\bibliographystyle{ACM-Reference-Format}
\bibliography{tzi,llm-bib}






\appendix
\input{appendix}
\end{document}

%% file: headers.tex
\usepackage[acronym]{glossaries}
\usepackage{amsmath,amsfonts, amsthm}
\usepackage{graphicx}
\usepackage{textcomp}
\usepackage{colortbl}
\usepackage{flushend} 
\usepackage{float}
\usepackage{subcaption}
\usepackage{multirow}
\usepackage{tabularx}
\usepackage{multicol}
\usepackage{xspace}
\usepackage{comment}
\usepackage{soul}
\usepackage{stfloats} 
\usepackage{longtable,booktabs}
\usepackage[version=4]{mhchem}
\usepackage{paralist}
\usepackage[compact]{titlesec}
\usepackage{setspace}
\usepackage[dvipsnames]{xcolor}
\usepackage{annotate-equations}
\usepackage{mathrsfs}
\usepackage{algorithm}
\usepackage{algpseudocode}
\usepackage{array}
\def\secformat{\bfseries}
\usepackage{listings}

\usepackage[most]{tcolorbox}
\newtcolorbox[auto counter]{finding}[1][]{%
    colback=blue!5,           
    colframe=blue!40,         
    boxrule=0pt,              
    leftrule=2mm,             
    sharp corners,            
    before upper={\textbf{Finding~\thetcbcounter:}~}, 
    fontupper=\normalfont,    
}

\definecolor{neongreen}{rgb}{0.0, 1.0, 0.0} 
\definecolor{neonpink}{rgb}{1.0, 0.07, 0.58} 
\sethlcolor{neonpink}

\usepackage{enumitem}
\definecolor{secblue}{HTML}{00008B}
\definecolor{subsecblue}{HTML}{0000FF}
\definecolor{subsubsecblue}{HTML}{0047AB}
\definecolor{captionblue}{HTML}{0000FF}
\definecolor{tableheader}{HTML}{203864}
\definecolor{paragraph}{HTML}{3B5E7F}
\definecolor{pgtext}{HTML}{0000CD}
\definecolor{hdrgray}{HTML}{BFCDDB}%
\definecolor{ltgray}{HTML}{DCDCDC}%
\definecolor{cellcolor}{HTML}{BFCDDB}

\titleformat{\subsection}{\secformat\color{secblue}}{\thesubsection}{0.5em}{}
\titleformat{\subsubsection}{\secformat\color{subsubsecblue}}{\thesubsubsection}{0.5em}{}

\setlist[itemize]{noitemsep, topsep=0pt, partopsep=0pt,left=0pt}

\titlespacing{\section}{0em}{0.0em}{0em}
\titlespacing{\subsection}{0em}{0.0em}{0em}
\titlespacing{\subsubsection}{0em}{0em}{0em}
\glsdisablehyper
\makeglossaries   
\newacronym{tl}{TL}{Transfer Learning}
\newacronym{hpc}{HPC}{High Performance Computing}
\newacronym{ml}{ML}{Machine Learning}
\newacronym{nlp}{NLP}{Natural Language Processing}
\newacronym{sipt}{M\textsubscript{$\mathcal{R}$}}{\texttt{Stacked Input Alignment Model}}
\newacronym{ipt}{M\textsubscript{$\mathcal{A}$}}{\texttt{Input Alignment Model}}

\newacronym{cn}{$\mathcal{M}odel\mathcal{X}$}{\texttt{Cross Prediction Model}}
\newacronym{sb}{SB}{Intel Sandy Bridge}
\newacronym{bgq}{BGQ}{IBM BG/Q}

\newacronym{mse}{MSE}{Mean Square Error}
\newacronym{nn}{NN}{Neural Network}
\newacronym{lp}{LP}{Linear Probing}
\newacronym{ft}{FT}{Fine-tuning}
\newacronym{ind}{IND}{In-Distribution}
\newacronym{ood}{OOD}{Out-of-Distribution}
\newacronym{fsl}{FSL}{Few-Shot Learning}
\newacronym{sssp}{SSSP}{Single-Source Shortest Path}
\newacronym{llms}{LLMs}{Large Language Models}
\newacronym{llm}{LLM}{Large Language Model}
\usepackage{tabularray}
\definecolor{sg}{HTML}{45B39D}

\definecolor{sgl}{HTML}{A2D9CE}
\definecolor{bl}{HTML}{5499C7}
\definecolor{bll}{HTML}{A9CCE3}
\definecolor{or}{HTML}{D35400}
\definecolor{orl}{HTML}{EDBB99}
\definecolor{usecase}{HTML}{008FAC}
\definecolor{deepskyblue}{HTML}{00BFFF}
\definecolor{purple}{HTML}{C4A9FF}
\definecolor{bb}{HTML}{DAE3F3}
\definecolor{oo}{HTML}{FBE5D6}
\definecolor{gry}{HTML}{EDEDED}
\definecolor{yl}{HTML}{FFC000}
\definecolor{yll}{HTML}{FFFFE0}

\newcommand{\tablecellbl}[1]{\cellcolor{bll}\textcolor{black}{#1}}
\newcommand{\tablecellor}[1]{\cellcolor{orl}\textcolor{black}{#1}}
\newcommand{\tablecellgr}[1]{\cellcolor{sgl}\textcolor{black}{#1}}

\newcommand{\tablecellhl}[1]{\cellcolor{blue!15}\textcolor{black}{#1}}

\usepackage{tikz}

\definecolor{green}{rgb}{0.1,0.1,0.1}

\newcommand{\xsbench}{\texttt{XSBench}\xspace}
\newcommand{\swlite}{\texttt{SW4Lite}\xspace}

\usepackage{diagbox}
\usepackage{fontawesome5}

\usepackage{caption}
\usepackage{dcolumn, array, booktabs}
\newcolumntype{.}{D{.}{.}{-1}}

\makeatletter

\newcommand{\scriptveryshortarrow}[1][3pt]{{%
    \hbox{\rule[\scriptratio\dimexpr\fontdimen22\textfont2-.2pt\relax]
               {\scriptratio\dimexpr#1\relax}{\scriptratio\dimexpr.4pt\relax}}%
   \mkern-4mu\hbox{\let\f@size\sf@size\usefont{U}{lasy}{m}{n}\symbol{41}}}}

\makeatother

%% file: short-abstract.tex
Large language models (LLMs) show promise for automated code optimization. However, without performance context, they struggle to produce correct and effective code transformations. Existing performance tools can identify bottlenecks but stop short of generating actionable code changes. Consequently, performance optimization continues to be a time-intensive and manual endeavor, typically undertaken only by experts with detailed architectural understanding.
To bridge this gap, we introduce \sysname, a modular, fully automated, end-to-end generative AI framework built on a multi-agent workflow. \sysname uses LLMs to map performance diagnostics from multiple reports to established, literature-backed code transformations, while unifying insight extraction, code generation, execution, and validation within a single pipeline. Across {$3{,}410$} real-world experiments on $10$ benchmarks and two \gls{hpc} mini-applications, \sysname generates $100\%$ correct code and improves performance in over $98.82\%$ of those experiments, achieving average gains of 8.02\%–79.09\% on NVIDIA GPUs.

%% file: intro-kdd.tex
As \gls{hpc} systems grow in complexity, writing high-performance GPU code increasingly requires deep architectural knowledge and expert interpretation of performance diagnostics. While profilers and analytical models can expose performance bottlenecks, translating those diagnostics into concrete and correct code changes remains a manual, expertise-driven process. In practice, few developers have routine access to vendor engineers or system architects, leaving most to sift through large volumes of diagnostic data without a clear path to actionable optimization.
Recent advances in \glspl{llm} for HPC code generation~\cite{roziere2023code,li2023starcoder,nichols2024hpccoder,nichols2024rlpf} suggest a possible alternative. However, existing approaches primarily emphasize syntactically valid code generation or rely on static analysis and prompt tuning~\cite{madaan2023learning,cummins2025llmcompiler}. As a result, they fail to incorporate dynamic performance signals---such as stall behavior, warp occupancy, and memory throughput---that dictate real-world performance. Our preliminary evaluation shows that directly providing GPU code to an LLM does not yield a single optimization that improve performance, indicating that syntax-level reasoning alone is insufficient for reliable LLM-based performance tuning.

This gap motivates our research question: \textit{Can multimodal runtime diagnostics be automatically mined to guide \glspl{llm} toward generating correct and performance-improving code?}
To answer this question, we introduce \textbf{\sysname}, an end-to-end framework that treats performance optimization as a data-driven reasoning problem.

Designing such a system presents four core challenges.
\textbf{First,} raw GPU telemetry can span hundreds of gigabytes per application, far exceeding the context limits of current models.
\sysname addresses this by extracting only the most salient signals from runtime profiles.
\textbf{Second,} performance diagnostics answer different questions and often point to different causes, making it unclear which signals to trust and how to act on them.
\sysname resolves this problem by explicitly mapping diagnostic patterns to application characteristics and linking them to well-defined optimization strategies.
\textbf{Third,} \glspl{llm} can generate code edits that compile but fail to improve performance because they do not target the actual bottleneck.
\sysname prevents this failure by explicitly requiring, through the prompt, that each code change cite the diagnostic signal it targets and be restricted to the code regions implicated by that signal.
\textbf{Fourth,} performance gains and correctness must be verified through execution.
\sysname closes the loop by coordinating model invocation, compilation, and bit-for-bit validation against reference outputs.

To quantify how well models reason over performance evidence, we introduce Evidence-Aligned Reasoning (EAR) metrics.
These model-agnostic metrics evaluate not only speedup, but also whether edits correctly cite diagnostic evidence, localize to identified hot spots, and preserve semantic correctness.
In this study, we use three complementary diagnostic sources—Roofline analysis~\cite{williams2009roofline}, PC stall sampling~\cite{zhou2021gpa}, and hardware event correlations~\cite{islam2016a}—which together capture what bottlenecks exist, how they manifest, and why they occur.

We evaluate \sysname using both microkernels and representative real-world applications across \textbf{3,410} controlled experiments.
Across all cases, diagnostic-informed prompting consistently outperforms baselines, yielding verifiable speedups and code transformations that align with microarchitectural ground truth.
We conduct all experiments on NVIDIA GPUs, which dominate both HPC and consumer computing ecosystems, ensuring that our optimization insights and conclusions generalize to real-world users rather than platform-specific edge cases.
In addition, we release an optimization corpus that aligns performance diagnostics, code transformations, and measured outcomes to support future research in data-driven and retrieval-augmented optimization.

To summarize, our contributions are:
\begin{itemize}
\item \textbf{A Multimodal Data Mining Pipeline:} A method to distill complex HPC performance logs into insights and a prompt template.
\item \textbf{Evidence-Aligned Reasoning (EAR) Metrics:} A novel set of metrics to quantify the consistency between diagnostic signals and generated code edits.
\item \textbf{Large-Scale Evaluation:} An analysis of 3,410 experiments on NVIDIA GPUs.
\item \textbf{The \sysname Framework:} A backend-LLM-agnostic, closed-loop system; will be made publicly available.
\item \textbf{Code Optimization Corpus:} A curated dataset of aligned performance diagnostics, code transformations, and measured outcomes to support future research; will be made publicly available.
\end{itemize}

By transforming a manual, expert-driven task into an automated workflow, \sysname democratizes access to performance engineering, serving as both an optimization framework and an educational tool for understanding the causal links between hardware diagnostics and code transformations.

%% file: approach3.tex
\begin{figure}[t]
    \centering
    \includegraphics[width=\columnwidth]{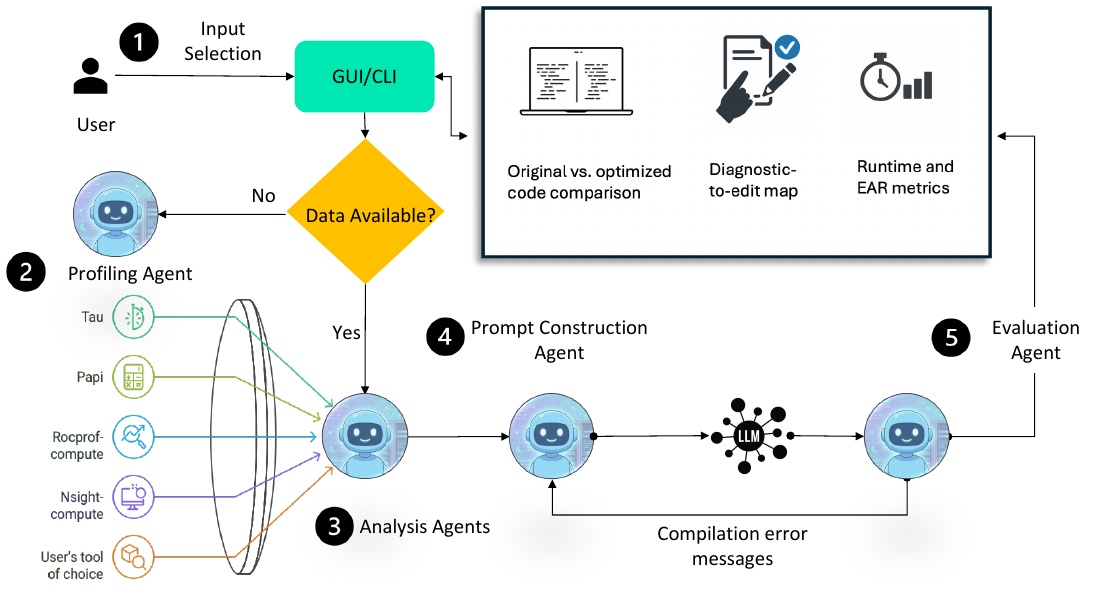}
    \caption{Overview of \sysname.}
    \label{fig:overview}
\end{figure}

Figure~\ref{fig:overview} shows the end-to-end workflow of \sysname.
Building on the analytical foundations described in this section, 
\sysname converts large, multi-source runtime profiles into validated, reproducible code transformations through the following steps:
(1)~Data Extraction and Insight Generation,
(2)~Prompt Construction and Optimization,
(3)~Evidence-Aligned Reasoning (EAR).
While this section focuses on the conceptual workflow, Section~\ref{sec:framework} later explains how these methodological steps are implemented as agents.

\subsection{Data Extraction and Insight Generation}
\label{sec:generate-insights}


\subsubsection{Automatically Identify Kernels Needing Optimization}
HPC applications often contain hundreds of kernels, and standard profiling tools report measurements for all of them.
However, users typically lack prior knowledge of which kernels are performance-critical, and LLMs impose strict token limits that make it infeasible to provide an entire application with all kernels and their associated measurements in a single request.
Moreover, only a small subset of kernels—typically 2--3—accounts for the majority (often $>$80--85\%) of total execution time, a long-observed property of HPC performance profiles~\cite{naderantahan2021topdown}.
These constraints require prioritizing only the most impactful kernels for downstream analysis.
To this end, \sysname automatically identifies performance-critical kernels using a lightweight profiling pass.
\sysname leverages the per-kernel execution time reported in the performance profiles to compute each kernel's fraction of the total application runtime, and selects the smallest kernel set $K^{*}$ such that
$K^{*} = \arg\min_{K} \left\{ \sum_{k \in K} T_k \ge \alpha \cdot T_{\text{total}} \right\}$. Here, $T_k$ denotes kernel runtime and $\alpha=0.8$ by default.
Only kernels in $K^{*}$ kernels are retained for detailed analysis and optimization, while all others are dropped. This step serves as a required preprocessing stage.

\subsubsection{Identify Program Characteristics from Roofline Analysis}
\label{sec:roofline_summarizing}
Roofline analysis provides a coarse-grained characterization of kernel efficiency by relating achieved throughput to arithmetic intensity and theoretical hardware limits~\cite{williams2009roofline,leinhauser2022amdroofline}. By quantifying arithmetic intensity (operations per byte), \sysname classifies kernels as either compute- or memory-bound.
Vendor tools generate performance diagnostics about achieved instruction throughput and memory bandwidth relative to hardware peaks. Utilizing vendor-specified hardware constants, \sysname computes utilization ratios $\rho \in [0, 1]$ for compute and memory resources. Based on the established HPC literature, we define a saturation threshold $\tau_{sat} = 0.70$~\cite{mess_memory_2024} resources are classified as \textit{underutilized} if $\rho < \tau_{sat}$ and \textit{saturated} if $\rho \ge \tau_{sat}$. These classifications are mapped to qualitative summaries. For instance, ``Compute underutilized (62\%), memory bandwidth saturated (91\%)". 

\subsubsection{Identify Salient Memory Access Stalls}
\label{sec:pc_summarizing}
While Roofline analysis explains \emph{what} resource bounds performance, it does not localize the root cause at the instruction or source-code level. When data are unavailable in the cache hierarchy, instructions cannot issue and execution stalls, indicating memory- or pipeline-related bottlenecks. To localize these effects, \sysname leverages PC Sampling method, which attributes stalled cycles to individual instructions and source lines~\cite{zhou2021gpa}. A high stall count therefore indicates that execution spends a significant fraction of time waiting rather than making forward progress.
Each stall type corresponds to a specific microarchitectural cause, such as memory barriers (\texttt{stall\_membar}), cache-level dependencies (\texttt{stall\_dependency}), or underutilized compute pipelines (\texttt{stall\_math}). The performance optimization literature has established optimization strategies for each. Some strategies are architecture-specific, while others generalize across GPU generations. For example, in Section~\ref{sec:posthoc-validation}, PC Sampling of \texttt{Accuracy} reveals dominant \texttt{stall\_membar} events on an NVIDIA GPU, motivating the use of \texttt{\_\_restrict\_\_} qualifiers for CUDA optimization.

Raw PC traces, however, can exceed hundreds of megabytes per kernel, making them unsuitable for direct LLM consumption. 
To address this challenge, \sysname applies an explicit saliency criterion that retains only stalls with significant execution impact. Specifically, \sysname (1) aggregates PC samples by source line and stall type; (2) selects, for each source line, the dominant stall that contributes at least $\tau\%$ of stalled cycles; and (3) emits a single textual summary linking the dominant stall to its corresponding \texttt{code snippet} and stalled-cycle fraction (e.g., \texttt{stall\_membar: 52\%}), discarding all other stalls. 
For instance, a 41-line \texttt{Accuracy} kernel generates 990\,MB of raw PC data, reducing it post-aggregation to 10\,MB and distilling it into a \texttt{<6}\,KB summary---a 99.9\% reduction. 
As many frequent stalls contribute negligibly to runtime and do not represent actionable bottlenecks.
Empirically, we find $\tau = 30\%$ sufficient as among the 18 available stall types, stalls contributing more than 30\% of total execution time dominate performance, and alleviating them yields measurable end-to-end speedups. 

\subsubsection{Identify Salient Application–Hardware Interactions}
\label{sec:importance}
\input{algo_counter_analysis}

Modern GPUs expose thousands of hardware performance counters, many of which are redundant or weakly correlated with execution time. To identify \emph{why} performance varies, \sysname formulates application--hardware interaction profiling as a sparse feature-selection problem. Let $N$ denote the number of profiling runs and $C$ the number of available counters. \sysname constructs a counter matrix $D \in \mathbb{R}^{N \times C}$ and a runtime vector $t \in \mathbb{R}^{N}$. To ensure comparability across heterogeneous metrics, each column of $D$ is z-score normalized to zero mean and unit variance.

Building on the \textbf{Ensemble Orthogonal Matching Pursuit (eOMP)} method in~\cite{islam2016a}, we develop a randomized greedy procedure (Algorithm~\ref{alg:selection}) to identify the most informative hardware counters. The counter-selection problem is inherently underdetermined ($C \gg N$). Dimensionality-reduction techniques such as PCA~\cite{ibrahim2016PCA} are unsuitable because they obscure the names of individual counters, which are required for providing context to the LLMs. Instead, we seek a sparse weight vector $a$ that minimizes $\|t - Da\|_2^2$ subject to $\|a\|_0 \leq \kappa$, where $\kappa \approx 5$ limits the selection to the most dominant performance signals.

As shown in Algorithm~\ref{alg:selection}, \sysname maintains a residual vector $r$ (initialized to $t$) and iteratively selects counters based on their correlation with the unexplained runtime, computed as $c_i = |D_i^T r|$. Rather than deterministically selecting the highest-correlated counter, the algorithm samples from a top-$\tau$ candidate pool, reducing sensitivity to noise and improving robustness. Counter importance is then aggregated across ensemble runs, yielding an average weight proportional to each counter’s ability to explain runtime variation.
The selected counters are mapped to natural-language descriptions (e.g., ``L2 cache miss rate'') using a vendor-derived dictionary~\cite{islam2019toward}. 
\subsection{Prompt Construction and Optimization}
\label{sec:prompt}
This stage combines the annotated kernel code with diagnostic summaries into a structured prompt consumable by the \gls{llm}. Each prompt includes the kernel, summaries produced in Section~\ref{sec:generate-insights}, and explicit guardrails that restrict edits to relevant code regions. These guardrails prohibit signature changes, duplicate kernels, and invalid syntax, and require the model to cite the diagnostic evidence motivating each optimization.
\sysname supports both commercial APIs such as GPT-5 and Gemini, as well as local backends such as Llama~3.1 via Ollama. Large inputs are automatically chunked, and model backends and parameters are configured through \texttt{config.yml} without modifying source code. Listing~\ref{lst:config} shows the prompt template used in this work.

\noindent\textbf{Listing 1:} The prompt template used by \texttt{\sysname}
\label{lst:config}
\lstdefinestyle{promptstyle}{
  basicstyle=\ttfamily\footnotesize,
  breaklines=true,
  columns=fullflexible,
  keepspaces=true,
  frame=single,
  escapeinside={(*@}{@*)},
  breakindent=0pt,
  breakautoindent=false,
  language={},
  keywords={}, 
  keywordstyle={}
}
\begin{lstlisting}[style=promptstyle]
# Begin Source Code
  <source_code_with_line_numbers>
# End Source Code
# Begin Performance Analysis
### (*@\textbf{STALL ANALYSIS:}@*)
<PC sampling analysis for exact lines of the source code>
### (*@\textbf{IMPORTANT HARDWARE COUNTERS:}@*)
<Lines describing the key hardware events/counter, what it means,
its impact score, and the kernel it is associated with>
### (*@\textbf{ROOFLINE ANALYSIS:}@*)
<Lines describing comments from roofline analysis using ncu>
# End Performance Analysis
# (*@\textbf{Instructions}@*)
You are an HPC performance optimization expert. You must optimize the source code provided above using only the performance data provided above.
Can you give me the fixed optimized code that will replace the full source code including headers and main function body? Consider the following instructions: 
    <Instructions describing common pitfalls and hallucination causes> 
    <Instructions describing the exact way to format and structure the output>
\end{lstlisting}


\subsection{Evidence-Aligned Reasoning (EAR)}
\label{sec:ear}
\input{ear.tex}

%% file: algo_counter_analysis.tex
\begin{algorithm}[t]
\caption{Important Counter Analysis}
\label{alg:selection}
\begin{algorithmic}[1]
\State \textbf{Initialize:} $r \gets t$, $S \gets \emptyset$, $\tau \gets 5$, $\kappa$, $E \gets 10$
\For{$e = 1$ \textbf{to} $E$}
    \While{$|S| < \kappa$}
        \State $c_i \gets |D_i^T r| \quad \forall i$ \Comment{Compute correlations}
        \State $\mathcal{I} \gets \text{top } \tau \text{ indices by } c_i$
        \State Sample $i \in \mathcal{I}$ randomly weighted by $c_i$
        \State $S \gets S \cup \{i\}$
        \State $r \gets r - D_S (D_S^T D_S)^{-1} D_S^T t$ \Comment{Update residual}
    \EndWhile
\EndFor
\State \textbf{Output:} Averaged selection ensemble
\end{algorithmic}
\end{algorithm}

%% file: ear.tex
To validate that \sysname's provided diagnostic information guided the responses of \gls{llms}, \sysname introduces a suite of five \emph{model-agnostic} metrics: 
the diagnostics, the edits made, and the post-hoc performance results.

\textbf{Evidence Coverage}
Evidence Coverage measures how often the model explicitly references or uses diagnostic clues in its generated edits. An edit is considered ``covered'' if its justification or location mentions at least one diagnostic item from the prompt 
(e.g., a stall type or counter name). 
High coverage implies that the \gls{llm}'s reasoning is successfully utilizing the provided evidence rather than hallucinating generic code changes, reported in Section~\ref{sec:results-ablation}.

\textbf{Localization Agreement.}
Localization Agreement assesses spatial alignment between the lines of code modified by the model and 
the regions identified as performance-critical in PC Sampling data. 
If edits occur near lines with high stall intensity, the model has localized the issue correctly. 
High agreement indicates that the LLM not only read the diagnostics but also acted on it.

\textbf{Directional Consistency}
Directional Consistency quantifies whether the applied optimizations produce runtime effects that match the direction implied by the diagnostics. 
For example, if the profiler indicates excessive memory stalls, a successful edit should reduce memory-related stall counters in post-optimization runs. 
We compare pre- and post-execution counter values and stall metrics (Figure~\ref{fig:accuracy_pcsampling_comparison}); 
the metric reports the proportion of edited kernels whose measured change aligns with the expected direction.
This validates that the model's ``fix'' corresponds to a real performance improvement rather than a coincidental change.

\subsection{Dataset Schema}
To ensure the reproducibility of our findings and support future benchmarking in code optimization, we will release our comprehensive dataset via an anonymous repository (\url{https://anonymous.4open.science/}). The corpus consists of $N=573$ unique records. Each record corresponds to a distinct tuple of \textit{(Application, Input Configuration, GPT-model)}. 
The dataset schema is partitioned into four primary categories: 
(1) \textbf{Contextual Metadata}, comprising the \textit{App Name}, the specific \textit{Prompt URL} used for LLM invocation, and the \textit{LLM} model identifier; 
(2) \textbf{Execution Environment}, detailing the \textit{Hardware Stack} and the \textit{Software Stack}; 
(3) \textbf{Optimization Specifics}, identifying the \textit{Applied Optimization} techniques, \textit{Ignored Optimization} candidates, and a direct \textit{Optimized Code URL}; and 
(4) \textbf{Performance Metrics}, recording \textit{Runtime}, \textit{Optimized Runtime}, and \textit{Error Notes}.
The URLs are local file paths in the repository. For full reproducibility of the binary generation and performance evaluation, each entry includes the \textbf{Compilation Command} and the \textbf{Execution Command} used within the specified environment. A representative instance of the dataset schema is provided below:
\texttt{\{App: ``Accuracy", Prompt: ``./prompts/.../accuracy/p\_ia.txt", LLM: ``GPT-5", HW: ``NVIDIA H100", SW: ``cuda/12.3,gcc/12.2", Compile: ``nvcc -O3 main.cu", Exec: "./a.out --size=1024", Opt\_Code: "./opt\_code/.../accuracy/opt\_ia.txt", Applied: [``Tiling"], Ignored: [``Vector"], Errors: ``", Base\_RT: ``12.4ms", Opt\_RT: ``8.2ms"\}}.

%% file: framework5-kdd.tex
\sysname implements a modular \textbf{multi-agent architecture} (Fig.~\ref{fig:overview}) to automate approximately 90\% of the optimization workflow. The pipeline executes in four stages:
(1) \textbf{Profiling:} If diagnostics are absent, the \texttt{Profiling Agent} invokes vendor tools to export raw JSON/CSV telemetry.
(2) \textbf{Analysis:} The \texttt{Analysis Agent} condenses raw logs into structured diagnostic summaries via the saliency and eOMP methods described in Sec.~\ref{sec:generate-insights} .
(3) \textbf{Prompting:} The \texttt{Prompt Construction Agent} assembles an optimized prompt containing the annotated kernel, diagnostic summaries, and relevant compiler feedback.
(4) \textbf{Evaluation:} The \texttt{Evaluation Agent} invokes the LLM, compiles the output, and executes each variant five times. It validates bit-level correctness and computes performance metrics, storing all artifacts in a versioned directory.
The system implements an iterative feedback loop: on compilation failure, the agent retries up to three times using error logs as feedback; persistent failures are surfaced to a dashboard for manual intervention. System parameters, including test suites and compiler flags, are managed via \texttt{config.yml}. 
Google-ADK is used to manage and maintain the agentic framework. 

\sysname provides both an interactive dashboard and a command-line interface. 
(1) The Streamlit dashboard (Figure~\ref{fig:optimas_ui} in the Appendix) allows users to select source code, diagnostics, and \gls{llm} settings, view optimized code with diagnostic annotations, inspect diagnostic-to-edit mappings, and compare runtime and EAR metrics; each run produces a timestamped results directory containing all artifacts. 
(2) In command-line mode, all parameters are specified in a single YAML file, and running \texttt{python run.py --config config.yml} executes the full workflow; each run is versioned with a UUID and a SHA-256 hash for reproducibility. 
(3) \sysname is modular. Hence, new profilers or LLM backends can be easily added without changing the other components.

%% file: setup2.tex
\subsection{Hardware and Environment}
We evaluate \sysname on an NVIDIA H100 GPU on an unnamed HPC cluster (hidden for double-blind review).  
Each system provides modern CPUs, high-bandwidth interconnects, and GPU-optimized software stacks.  
Experiments use CUDA~12.3 and Nsight Compute~2025.1.1.0.  
All CUDA sources are compiled with \texttt{-lineinfo} for PC sampling to enable address–source mapping.  
For runtime measurements, \texttt{-lineinfo} and \texttt{-G} are disabled to avoid any performance impact.  
For LLMs, \gpt\ and \gemini\ are accessed via APIs, while \llama\ is deployed locally using Ollama.  

\subsection{Benchmarks and Configurations}

\input{table5}
We evaluate \sysname on a diverse suite of 10 GPU applications comprising 16 distinct kernels, selected to span a broad spectrum of arithmetic intensities and memory access patterns. As detailed in Table~\ref{tab:app-config-summary}, the suite is categorized into two tiers. 
(1) We utilize fundamental microbenchmarks to isolate specific architectural bottlenecks. These include bandwidth-bound primitives (\texttt{BabelStream}~\cite{deakin2018babelstream}), shared-memory intensive patterns (\shm), and specialized kernels targeting compute-bound (\acc, \extrema) and irregular memory-bound (\sobol, \adjacent) behaviors.
(2) To assess the efficacy on production-grade workloads, we include two Department of Energy (DOE) proxy applications: \xsbench~\cite{tramm2014xsbench} and \swlite~\cite{wu2021performance}. These applications capture complex HPC motifs, specifically the irregular memory lookups of Monte Carlo neutron transport (\xsbench) and the structured stencils of seismic wave propagation (\swlite). For the multi-kernel \swlite application, we target the five most computationally significant hot-spots. 



\subsection{Data Collection Tools}
\textbf{Roofline Analysis.}
On NVIDIA GPUs, we use Nsight Compute (\texttt{ncu}) to extract arithmetic intensity, achieved throughput, and diagnostic summaries using
\textcolor{blue}{\texttt{ncu ----csv ----import <output\_file> ----page details > <roofline\_file>.csv}}.
\textbf{PC Sampling.}
We collect PC sampling data using Nsight Compute's replay mode, with \texttt{-lineinfo} enabled to map sampled instructions back to source lines.
\textbf{Hardware Performance Counters.}
We collect the full set of available hardware performance counters ($1200+$) using Nsight Compute's \texttt{full} preset:
\textcolor{blue}{\texttt{ncu ----target-processes all ----set full -o <output\_file> ./<app> <args>}}.
Since raw counter names are often cryptic, we augment each selected counter with its semantic description in the prompt. Using Algorithm~\ref{alg:selection} for counter reduction, we empirically observe that increasing $\kappa$ beyond $5$ does not change the generated responses. So, for all experiments in this work, we set $\kappa=5$.
Prior work shows that contextualized diagnostics significantly improve LLM reasoning effectiveness while reducing inference overhead~\cite{banday2025role}. Counter descriptions and mappings are drawn from the Dashing repository~\cite{islam2019toward}.

\subsection{Baselines}
\label{sec:baselines}
We compare \sysname's optimizations against the original unoptimized code under seven diagnostic configurations:  
(1)~PC, (2)~IA (counter-based interaction analysis), (3)~Roofline,  
(4)~PC+IA, (5)~PC+Roofline, (6)~IA+Roofline, and (7)~PC+IA+Roofline.  
These seven scenarios correspond directly to the labels used in all ablation figures.  
To demonstrate applicability to \gls{hpc} workloads, we also evaluate \xsbench~\cite{tramm2014xsbench} and \swlite~\cite{wu2021performance}.  
For \swlite, we optimize both the kernel \texttt{addsgd4\_SM} and the five most time-consuming kernels (forcing\_dev, rhs4\_X<0, 0>,rhs4\_Y<0, 1> and rhs4\_Y<0, 0>) in the suite.  
Hand-tuned expert baselines are excluded, as they depend on domain knowledge and are not publicly available for us to obtain.  
We also tested ``code-only" baselines with no diagnostics at the same temperature and found that it yielded no valid optimizations. This observation provides the lowest baseline of $0\%$.  


\subsection{LLM Reproducibility vs.\ Creativity}
\label{sec:temapareture}
The temperature parameter controls the trade-off between reproducibility and creativity.  
Higher values encourage diverse transformations but reduce consistency; lower values yield reproducible yet narrower changes.  
We empirically tuned the sampling temperature across the range $[0, 1]$ and observed that $\texttt{temperature}=0.15$ yields the optimal trade-off between deterministic reproducibility and generative diversity. Consequently, we fix this value for all subsequent experiments.
However, users can configure this parameter through the \texttt{config.yml} of \sysname.  
Related work for temperature tuning (e.g., TURN~\cite{du2025turn}) can be leveraged to automatically adjust diversity based on model entropy.

%% file: table5.tex
\begin{table}[t]
\scriptsize 
\centering
\caption{Benchmark applications, their size in line of code (LOC), and tunable configuration parameters.}
\label{tab:app-config-summary}

\begin{tabularx}{\columnwidth}{|p{1.2cm}|c|X|}
\hline
\textbf{App.} & \textbf{LOC} & \textbf{Configuration Description} \\
\hline

\texttt{Copy} & 2 & \multirow{5}{=}{
    \texttt{ARRAY\_SIZE}$\in$[2000896, 4000768, 6000640], \newline
    \texttt{num\_times}$\in$[100, 300], step=100; \newline
    Default=(2000896, 100). \newline
    \texttt{ARRAY\_SIZE}: dataset size; \newline
    \texttt{num\_times}: repetition count.
} \\
\cline{1-2}
\texttt{Mul} & 3 & \\ 
\cline{1-2}
\texttt{Add} & 2 & \\
\cline{1-2}
\texttt{Triad} & 3 & \\
\cline{1-2}
\texttt{Dot} & 19 & \\
\hline

Shmemench~\cite{meswani2012tools} & 35 &
\texttt{rep}$\in$[300, 1500], step=300; Default=600. \newline
\texttt{rep}: kernel repetition count. \\
\hline

\acc~\cite{jin2023a} & 41 &
(\texttt{nrows}, \texttt{ndims}, \texttt{top\_k}, \texttt{rep}) = $8192 \times [5000, 10000] \times [10, 20] \times [100, 300]$; \newline
Default=(8192, 5000, 10, 100). \newline
\texttt{nrows}: prediction vectors; \texttt{ndims}: dimensions/vector; \texttt{top\_k}: threshold; \texttt{rep}: repetition count. \\
\hline

\sobol~\cite{jin2023a} & 195 &
(\texttt{n\_vectors}, \texttt{n\_dimensions}, \texttt{repeat}) = $[10\text{K}, 1\text{M}] \times [1\text{K}, 10\text{K}] \times 100$; \newline
Default=(10K, 1K, 100). \newline
\texttt{n\_vectors}: vectors generated; \texttt{n\_dimensions}: dimensions/vector; \texttt{repeat}: repetition count. \\
\hline

\adjacent~\cite{nvidia_cccl_adjacent_difference} & 180 &
(\texttt{repeat})=[10, 1000]; Default=100. \newline
\texttt{repeat}: number of repetitions of adjacency test. \\
\hline

\extrema~\cite{Thompson2021_cuSignal} & 160 &
(\texttt{timesteps}, \texttt{paths}, \texttt{runs}, $T, K, r, \sigma$) varied; \newline
Default=(100, 32, 1, 1.0, 4.0, 0.06, 0.20). \newline
\texttt{timesteps}: simulation timesteps; \texttt{paths}: Monte Carlo paths; \texttt{runs}: independent runs; $T$: maturity; $K$: strike price; $r$: interest rate; $\sigma$: volatility. \\
\hline

\texttt{addsgd4\_SM}~\cite{wu2021performance} & 149 &
(\texttt{input\_datasets}) = [pointsource, uni, uni\_rev, skinny, gaussianHill]; \newline
Default=(pointsource). \\
\hline

\xsbench~\cite{tramm2014xsbench} & 81 &
(\texttt{s}, \texttt{G})=[small, large] $\times$ [unionized, nuclide, hash]; \newline
Default=(large, unionized). \newline
\texttt{s}: problem size; \texttt{G}: grid search type. \\
\hline
\swlite~\cite{wu2021performance} Application. Many kernels. Top 4 are selected. & 14017 &
(\texttt{input\_datasets}) = [pointsource, uni, uni\_rev, skinny, gaussianHill]; Default=(pointsource). 
\textbf{\swlite exemplifies multi-kernel dependencies.} We focused on the top five kernels (including \texttt{addsgd4\_SM}), which generate 80\% of the execution time out of the 61 total.
\\
\hline

\end{tabularx}
\end{table}

%% file: results3.tex
\input{table_ablation_study_all_llms}
In this section, we evaluate the effectiveness of \sysname for 10 benchmarks using average wall-clock runtime improvement, computed over 10 runs.
Correctness is verified via bit-for-bit comparison against reference outputs, and all LLM-generated variants compile and pass functional tests under deterministic prompting.
Specifically, we: (i)~\textbf{select} an open-source LLM, (ii)~\textbf{compare} code generation models, (iii)~\textbf{quantify} the contribution of each insight source via ablation and explain insight effectiveness across application characteristics, (iv)~\textbf{evaluate} optimizations across configurations, (v)~\textbf{validate} bottleneck fixes via post-hoc analysis, (vii)~\textbf{demonstrate} on \gls{hpc} proxy applications, and (viii)~\textbf{quantify} \sysname's overhead.
Table~\ref{tab:default-opt-summary} in the Appendix summarizes the specific optimizations suggested for each applications. We restrict this analysis to the default configuration (Table~\ref{tab:app-config-summary}) for each benchmark unless otherwise specified to keep the experiment space tractable.  

\subsection{Open-Source LLM Selection}
\label{sec:llms}
\input{llms}

\subsection{Comparison of Different Code Generation Models}
\label{sec:results-ablation}
\input{results-all-llms}


\subsection{Effectiveness of Insight Types by Application Characteristics}
\label{sec:results-app-characteristics}
\input{results_ablation_and_application_class}
\subsection{Evaluation the effectiveness of optimizations across many configurations}
\label{sec:allconfig}
\begin{figure}[t]
    \centering
    \includegraphics[width=\columnwidth]{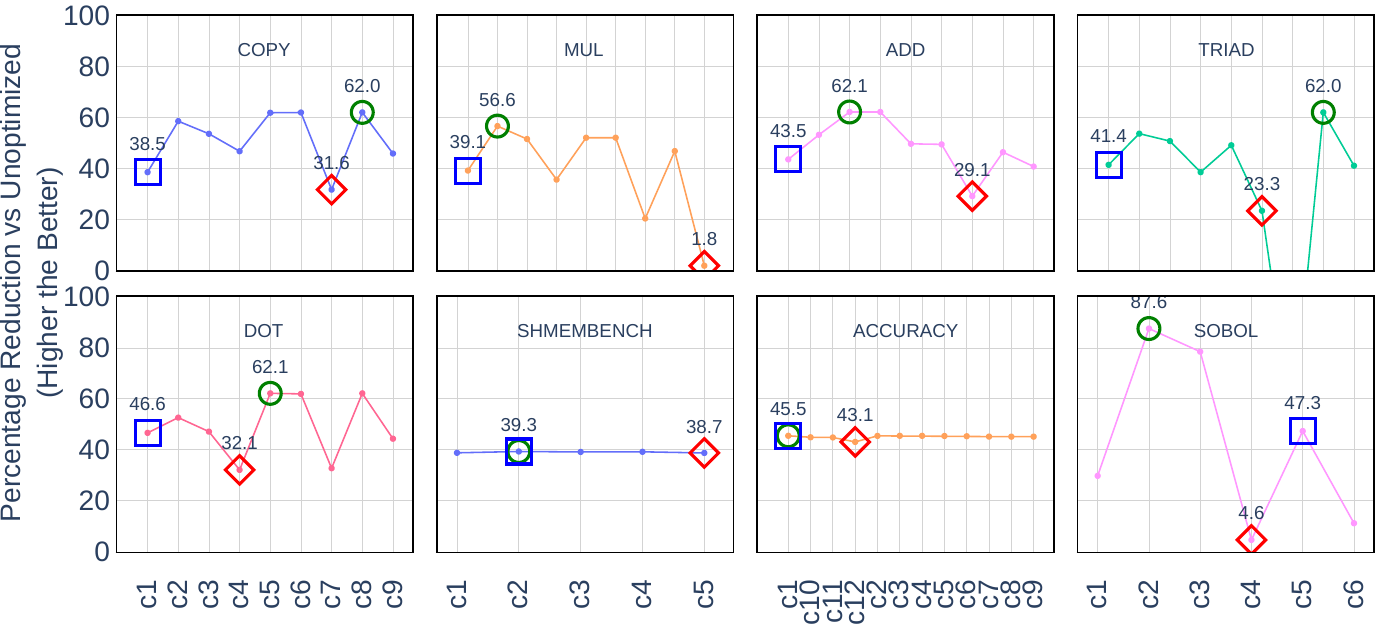}
    \caption{%
    Performance improvements across input configurations. 
    Blue rectangles represent improvements for default configuration (Table~\ref{tab:app-config-summary}), green circles indicate maximum improvement, and red diamonds indicate minimum improvement. This figure shows that optimizations generated for one configuration also yield performance improvements across other configurations, however, the margin varies across applications. \acc achieves the most stable gains (43.1\%–45.5\%), while that for \sobol varies widely (4.6\%–87.6\%).
    }
    \label{fig:allapps-allconfig}
\end{figure}
The objective of this experiment is to assess whether \sysname's code transformation decided based on a single configuration generalize across diverse input sizes and configurations. 
Each run compares the unoptimized version to the \sysname-optimized version generated using analytics from the default configuration. 
Figure~\ref{fig:allapps-allconfig} plots the percentage reduction in execution time: the X-axis shows each configuration; the Y-axis shows the percentage reduction in runtime relative to the unoptimized baseline.

\textbf{Key observations.} From Figure~\ref{fig:allapps-allconfig}, we observe that: 
(1) execution-time reductions persist across all but 1 out of 68 configurations spanning 8 applications, and 
(2) the only configuration of \triad that shows no improvement corresponds to an array size of 6{,}000{,}640 with 100 kernel repetitions. 
Further investigation reveals that this configuration shifts the kernel into a different bottleneck regime, where the dominant cost moves from memory throughput to loop overhead and launch latency, reducing the effectiveness of the applied memory-focused optimization.
(3) Kernels like \shm (up to 69.10\%) and \sobol (up to 87.60\%) see the largest gains, due to recurring bottlenecks across input ranges that let~\sysname generalize effectively. However, the improvements for \sobol drop for $C4$, indicating an input-sensitive shift in the dominant bottleneck from compute and instruction throughput to memory access and control divergence caused by changes in stride and loop structure under this configuration. (4) In contrast, \acc consistently improves (43.1\%–45.5\%) across all inputs. The applied transformation (\texttt{\_\_restrict\_\_}) targets aliasing-related stalls that persist regardless of scale, indicating stable bottlenecks. These observations validate that optimizations from one configuration can transfer when performance characteristics remain consistent.

\subsection{Validate using Post-hoc Reasoning}
\label{sec:posthoc-validation}
\begin{figure}[t]
    \centering    
    \includegraphics[width=\columnwidth]{Figures/pcsample_comp.pdf}
\caption{%
Comparison of stall occurrences for the \acc kernel before and after optimization. 
The Y-axis lists source line numbers with shortened code snippets, and each bar is labeled by its dominant stall type. At line 28, \texttt{stall\_wait} occurrences decrease from $45{,}387 \rightarrow 22{,}642$, with similarly large reductions at lines 29--31, confirming that the applied optimizations directly reduce dominant stall sources.}
    \label{fig:accuracy_pcsampling_comparison}
\end{figure}
This experiment assesses whether \sysname's code transformations mitigate the performance bottlenecks identified during analysis by re-profiling the \acc kernel before and after optimization using its default configuration. 
Performance profiling revealed poor warp readiness (0.16 eligible warps per cycle), inefficient DRAM fetch (2.7 sectors per 128-byte cache line), and persistent memory stalls, including \texttt{stall\_barrier} and \texttt{stall\_membar}. To address these, \sysname added the \texttt{\_\_restrict\_\_} keyword to \texttt{Xdata} and \texttt{labelData} (Table ~\ref{tab:default-opt-summary}). This keyword disambiguates pointer aliasing, which makes the compiler safely reorder memory accesses and apply coalescing optimizations. Published literature confirms that this keyword 
is used to enable contiguous memory access, which reduces transactional overhead~\cite{cuda_guide2024,williams2009roofline,crago2018exposing}. The post-optimization metrics show that (1) scheduler efficiency improves from one instruction every 6.4 cycles to 3.5, (2) eligible warp rate increases from 0.16 to 0.40, and (3) DRAM fetch efficiency improves as indicated by a lower L2 miss-to-sector ratio. These improvements confirm that the \gls{llm} correctly identified a transformation that targets a bottleneck and aligns with expert practices.

\subsection{Demonstration on HPC Proxy Applications}
This experiment studies whether \sysname's benefit applies to large and complex real-world \gls{hpc} applications: \xsbench and \swlite. 
%
Both \xsbench and \swlite achieve substantial performance gains—0–11.58\% and 13.15–14.56\%, respectively---when leveraging all available diagnostic sources. These workloads represent high-complexity regimes: \xsbench is dominated by a single long, intricate kernel, while \swlite exhibits performance dependencies across multiple interacting kernels.
In such settings, no single diagnostic is sufficient to capture all performance limiters. By combining complementary sources, \sysname captures heterogeneous inefficiencies, including instruction-level limitations in long-running kernels and interaction overheads across multi-kernel execution flows. This multi-source strategy effectively serves as a safety net, providing comprehensive bottleneck coverage where narrower analyses would likely miss critical optimization opportunities.

\subsection{Cost and Overhead Accounting}
\label{sec:costs}

\textbf{API usage and cost.}
For \gpt, 773{,}000 input and 910{,}000 output tokens result in a cost of \$9.83 (average \$0.085 per experiment).
For \gemini, 954{,}846 input and 901{,}687 output tokens result in a cost of \$16.54 (average \$0.1476 per experiment). 

\textbf{Overhead.} 
HPC users commonly spend hours to days collecting performance data~\cite{RooflineScalingTrajectories,lee2020roofline,williams2009roofline,amd_roofline_micro2022,lawson2015experimentationprocedureoffloadedminiapps,zhou2021gpa,zhou2020tools,islam2016a,dashing,islam2019toward,zaeed2024characterize,islam2025data,banerjee2016cmtbone}. 
Despite this cost, they cannot systematically translate these results into concrete code optimizations. 
\sysname addresses this gap using the data users already collect. 
Consequently, profiling time dominates the end-to-end workflow, not \sysname. 
For example, profiling \swlite required $\approx$16\--24 hours, while \sysname completed one optimization iteration in $<10$ minutes: 
(i)~$\le$30\,s for summarization, 
(ii)~2 minutes on average for LLM inference for commercial LLMs and 4 minutes on average for LLama per iteration, and 
(iii)~4 minutes on average for compilation and validation, so in total optimization done in a few minutes which is $<1.5\%$ of the data collection time.

\subsection{Discussions}
The utility of performance insights depends on the dominant bottlenecks of a given application. Memory-bound kernels often benefit from lightweight analyses such as Roofline, whereas complex, multi-kernel applications require the complementary use of multiple diagnostics. This observation confirms that multi-source prompting is most effective for real-world applications that are sure to have multi-kernel dependencies, especially when it is producing a speedup 3,370 times out of the 3,410 experiments.
While \sysname currently provides a principled alternative to costly exhaustive configuration search, we view it as a foundational baseline for further automation. Future work will explore a closed-loop optimization workflow in which an evaluation agent autonomously triggers and re-executes diagnostics based on observed performance outcomes. At present, this feedback loop is user-driven, requiring manual re-invocation of \sysname's profiling agents on optimized code.

%% file: table_ablation_study_all_llms.tex
\begin{table}[!t]
\footnotesize
\caption{Ablation study of diagnostic sources on NVIDIA CUDA kernels, reporting runtime improvement (\%) across diagnostic combinations and LLM backends. Higher is better.
Across 10 runs, variation in performance \emph{improvements} (not run-to-run noise) is within $\pm10\%$ on average, higher for short- and lower for long-running kernels; hence, variance details are omitted for readability.
}
\label{tab:ablation-multisource}
\resizebox{\columnwidth}{!}{
\begin{tabular}{|l|c|c|c|c|c|c|c|}
\hline
\textbf{Dataset} & \textbf{PC} & \textbf{IA} & \textbf{Roofline} & \textbf{PC+IA} & \textbf{PC+Roofline} & \textbf{IA+Roofline} & \textbf{PC+IA+Roofline} \\
\hline
\multicolumn{8}{|c|}{\tablecellbl{\textbf{GPT-5}}} \\ \hline
Copy        & 33.33 & \tablecellhl{39.58} & 35.42 & 35.42 & 35.42 & 35.42 & 35.42 \\
Mul         & 32.65 & \tablecellhl{38.78} & 34.69 & 34.69 & 34.69 & 34.69 & 34.69 \\
Add         & 30.30 & \tablecellhl{33.33} & 31.82 & 31.82 & 31.82 & 31.82 & 31.82 \\
Triad       & 31.34 & \tablecellhl{34.33} & 32.84 & 32.84 & 31.34 & 32.84 & 32.99 \\
Dot         & 34.69 & 26.53 & 36.73 & 36.73 & 36.73 & 36.73 & \tablecellhl{36.73} \\
Sobol       & \tablecellhl{41.60} & \textcolor{red}{-4.47} & 40.39 & 40.38 & 3.81  & 40.00 & 38.43 \\
Accuracy    & 78.29 & 77.90 & 78.29 & 78.46 & \tablecellhl{79.09} & 78.68 & 78.42 \\
Shmembench  & 42.00 & 64.93 & 42.13 & \tablecellhl{64.93} & 42.13 & 36.27 & 36.27 \\
Adjacent    & 51.03 & 51.00 & 51.02 & \tablecellhl{56.40} & 51.00 & 54.05 & 51.00 \\
Extrema     & 16.36 & 28.99 & 18.87 & 32.71 & 17.74 & 23.48 & \tablecellhl{41.20} \\
\hline
\multicolumn{8}{|c|}{\tablecellgr{\textbf{Gemini-2.5}}} \\ \hline
Copy        & 33.33 & 33.33 & 35.42 & 35.42 & 35.42 & 35.42 & \tablecellhl{35.42} \\
Mul         & 32.65 & 32.65 & 34.69 & 34.69 & 34.69 & 34.69 & 34.69 \\
Add         & 31.82 & \tablecellhl{33.33} & 31.82 & 31.82 & 31.82 & 31.82 & 31.82 \\
Triad       & 31.34 & \tablecellhl{34.33} & 32.84 & 32.84 & 32.84 & 32.84 & 32.99 \\
Dot         & 34.69 & 26.53 & 36.73 & 36.73 & 36.73 & 36.73 & \tablecellhl{36.73} \\
Sobol       & 77.10 & \tablecellhl{79.83} & 76.04 & 77.08 & 72.98 & 74.31 & 76.05 \\
Accuracy    & 0.25  & 0.12  & 0.28  & -0.66 & 0.20  & 0.26  & \tablecellhl{0.43} \\
Shmembench  & 51.53 & \tablecellhl{51.65} & 14.73 & 14.36 & 51.61 & 49.83 & 15.07 \\
Adjacent    & 51.11 & 50.69 & 50.59 & 29.61 & 49.56 & 51.11 & \tablecellhl{51.11} \\
Extrema     & 4.47  & 2.23  & 0.72  & \tablecellhl{26.57} & 1.84  & 19.78 & 9.09 \\
\hline
\multicolumn{8}{|c|}{\tablecellor{\textbf{LLaMA}}} \\ \hline
Copy        & 60.42 & 58.33 & 60.42 & 56.25 & 60.42 & 58.33 & \tablecellhl{62.50} \\
Mul         & 61.22 & 59.18 & \tablecellhl{61.22} & 57.14 & 59.18 & 46.94 & 59.18 \\
Add         & 60.61 & 60.61 & 60.61 & 60.61 & 60.61 & 59.09 & \tablecellhl{60.61} \\
Triad       & 61.19 & 59.70 & 59.70 & 58.21 & 61.19 & 59.70 & \tablecellhl{61.19} \\
Dot         & 20.41 & 18.37 & 18.37 & 16.33 & \tablecellhl{20.41} & 18.37 & 18.37 \\
Sobol       & 18.25 & 18.06 & 18.27 & 17.97 & 11.08 & \tablecellhl{18.25} & 17.73 \\
Accuracy    & 0.25  & 0.12  & 0.28  & -0.66 & 0.20  & 0.26  & \tablecellhl{0.43} \\
Shmembench  & 18.38 & 26.80 & 24.55 & 29.87 & 18.39 & \tablecellhl{30.99} & 18.40 \\
Adjacent    & 51.03 & 51.00 & 51.02 & \tablecellhl{69.69} & 51.00 & 50.09 & 51.00 \\
Extrema     & 8.65  & 8.65  & \tablecellhl{34.65} & 8.65  & 8.60  & 34.34 & 8.63 \\
\hline
\end{tabular}
}
\end{table}

%% file: llms.tex
The objective of this experiment is to identify an open-source \gls{llm} that can perform code optimizations comparable to commercial models.  
We evaluate several widely used open-source \glspl{llm} known for coding tasks, namely, \texttt{Mistral}, \texttt{Qwen2.5:32B}, \texttt{Gemma:8B}, and \texttt{Gemma:32B}.  

\textbf{Key observations.}
Our evaluation reveals recurring limitations that make all evaluated open-source models except \llama unsuitable for integration into \sysname.
(1) Models such as \texttt{Mistral}, \texttt{Qwen2.5}, and \texttt{Gemma:8B} fail to translate performance diagnostics into actionable optimizations. While \texttt{Mistral} and \texttt{Qwen2.5} produce generic examples or paraphrased snippets rather than modifying the provided source code, \texttt{Gemma:8B} fails to ground diagnostic evidence (e.g., stalls) in specific source locations. This behavior indicates a collective inability to align diagnostics to actionable code edits. In contrast, \llama identifies mapping diagnostic signals to specific code constructs by leveraging its superior instruction-following capabilities and reasoning coherence.
(3) While \texttt{Gemma:32B} follows partial instructions, it frequently violates critical guardrails, such as preserving kernel signatures or maintaining code structure, producing edits that are unsafe or non-deployable. \llama consistently respects these constraints under deterministic prompting, because its larger parameter count provides the necessary contextual capacity to simultaneously maintain global constraints while performing local optimizations, avoiding the attention drift common in smaller architectures.
For this reason, we select \llama as the open-source baseline for all remaining experiments.

%% file: results-all-llms.tex
The objective of this experiment is to evaluate how different \glspl{llm} optimize the same GPU kernels when provided with identical prompts and performance insights.
We evaluate three \glspl{llm}: \gpt, \gemini, and \llama to compare the quality of optimization suggestions by measuring their resulting runtime improvements.
Table~\ref{tab:ablation-multisource} summarizes all results an NVIDIA~H100 GPU.  

\textbf{Key observations.}
Across all benchmarks, the global average evidence coverage is 73.3\%, indicating that most edits are explicitly driven by performance indicators. \gpt{} achieves the highest evidence coverage (83.1\%) and the largest average speedups, particularly on multi-kernel applications (\acc, \sobol, \swlite), reflecting high reasoning completeness and effective aggregation of multiple diagnostics. However, on single-bottleneck streaming kernels (\add, \triad), \gpt{} tends to over-optimize by applying additive transformations beyond saturation, revealing a lack of negative reasoning—i.e., insufficient pruning of marginally beneficial edits.
\gemini{} exhibits similar behavior with slightly lower coverage (75.6\%) and higher run-to-run variance. It performs comparably on large kernels but underperforms on stall-sensitive kernels (\shm), consistent with partial loss of fine-grained diagnostic context due to prompt-length constraints.
\llama{} shows lower evidence coverage (49.2\%) and smaller but stable gains. Its optimizations are conservative and generic, reflecting limited exposure to GPU-specific idioms and weaker cross-source reasoning. While less aggressive, \llama{} remains effective under cost, privacy, and locality constraints, demonstrating that \sysname's gains do not depend on proprietary models.

We further evaluate model reliability using our EAR metrics. On average across all applications, \gpt implements 23 optimizations while withholding only 5 suggestions and producing zero hallucinations (instances where the text claims an edit that is absent in the code). In comparison, \gemini exhibits a more exploratory but less consistent behavior, implementing 14.5 optimizations and withholding 6, but suffering from 1 hallucination on average. In contrast, \llama adopts a highly conservative strategy, implementing only 5.5 optimizations with zero hallucinations, prioritizing precision and safety over optimization volume. In terms of localization agreement, \gpt applies 100\% optimizations at the correct code locations, compared to 85\% for \gemini and 80\% for \llama.

%% file: results_ablation_and_application_class.tex
The objective of this experiment is to conduct an ablation study that isolates the contribution of each performance insight source, both individually and in combination, and to evaluate how their effectiveness depends on application characteristics.
To ensure experimental tractability and avoid confounding effects from configuration changes, we fix the default configuration and vary only the diagnostic sources provided to the LLMs. 
This study serves two purposes: (i) to quantify how different insights guide LLM, and (ii) to distill generalizable knowledge about which sources are most effective for different classes of HPC applications.

\textbf{Key observations.}
From Table~\ref{tab:ablation-multisource}, we can observe the following.
(1) Each diagnostic source independently yields measurable performance improvements across most kernels, with gains ranging from \textbf{3.81--79.09\%} for \gpt, \textbf{0.12--79.83\%} for \gemini, and \textbf{0.1--69.69\%} for \llama. This observation confirms our hypothesis that even a single performance insight provides a measurable benefit over using no diagnostic guidance.
(2) Different performance insight sources expose distinct classes of bottlenecks and are therefore effective in different regimes.
Roofline analysis and PC Sampling provide \emph{structural signals} that reveal whether performance is limited by memory bandwidth, instruction throughput, synchronization, or dependency-driven stalls, making them especially actionable for memory- or stall-bound kernels.
This is evident in the \texttt{sobol} kernel, where Roofline analysis exposes severe underutilization—low achieved occupancy (25.6\% vs.\ 100\% theoretical) and poor issue efficiency (one instruction every 1.9 cycles)—while PC Sampling localizes persistent stalls such as \texttt{stall\_wait}, \texttt{stall\_branch\_resolving}, and \texttt{stall\_membar}, indicating dependency-driven short-scoreboard effects. In contrast, IA surfaces only a small number of moderately weighted counters, primarily related to DRAM and L1TEX writes, which lack the line-level specificity needed to guide structural transformations.
Guided by Roofline and PC Sampling, \sysname applies expert-aligned optimizations—raising occupancy via \texttt{\_\_launch\_bounds\_\_(64)}, precomputing stride-related values to eliminate repeated \texttt{\_\_ffs} calls, and unrolling loops to increase instruction-level parallelism (Table~\ref{tab:default-opt-summary}).
(3) Dual-source combinations provide improvements (\textbf{1.0--79.09\%} for \gpt, \textbf{0.25--69.69\%} for \llama, and \textbf{0.11--78.3\%} for \gemini) when they provide complementary yet non-conflicting pieces of information about the bottlenecks. {For instance, Roofline analysis of \acc kernels shows us that there is a large scoreboard dependency on L1TEX operations, and PC sampling data shows exactly in which line stalls like \texttt{stall\_long\_sb\_ratio} and \texttt{stall\_wait\_ratio} dominate to pinpoint this information.}
(4) The full combination of all three sources (\texttt{PC+IA+Roofline}) delivers gains for multi-kernel applications (\acc, \sobol, \swlite, \xsbench) that exhibit mixed bottlenecks across kernels and execution phases and therefore benefit most from combining signals, with average improvements spanning \textbf{31.8--78.98\%} for \gpt and comparable ranges for \llama and \gemini.

\textbf{Why combining diagnostic sources is not monotonic.}
The benefit of combining diagnostic sources is not monitonic because:
(1) While memory stalls listed in PC samples may require localized instruction refactoring, a broader data-movement inefficiency identified by Roofline analysis may require a larger structural change. In this context, safety guardrails prevent unsafe rewrites by blocking aggressive transformations, unless multiple sources jointly justify a deeper change.
(2) Adding more sources increases prompt length, which can dilute attention to the dominant bottleneck signal. As a result, we interpret multi-source prompting as a robustness mechanism that increases the likelihood of identifying a correct optimization path, rather than as a strictly additive signal.

%% file: related-work.tex
Performance analysis methodologies such as Roofline modeling~\cite{williams2009roofline,RooflineScalingTrajectories,lee2020roofline,amd_roofline_micro2022,Ilic14CRM}, PC Sampling~\cite{zhou2021gpa,chabbi2013effective,zhou2020tools,zhou2021measurement} and hardware performance counters~\cite{uhsadel2008exploiting, islam2019dashing,islam2016a} help identify architectural inefficiencies (e.g., memory bottlenecks, TLB misses) but require manual interpretation and expert-guided code tuning. Similarly, PC Sampling tools such as Nsight~\cite{nvidia2023nsight,zhou2021gpa} localize stalls to source lines but leave the optimization step to the user. While these approaches are effective at diagnosis, they do not automate the path to resolution. \sysname provides an extendable framework that complements this ecosystem by structurally translating these diagnostics into insights that can guide \gls{llms} to generate actionable code changes, thus closing the loop between profiling and code transformation.

General-purpose models such as Codex~\cite{chen2021evaluatinglargelanguagemodels}, StarCoder~\cite{li2023starcoder}, and CodeLlama~\cite{roziere2023code} are trained for code completion and refactoring but lack awareness of runtime performance bottlenecks. Similarly, Star-Agents~\cite{zhou2024star} learn from version histories but do not incorporate profiling data. As a result, these models tend to propose generic transformations that are disconnected from actual hardware constraints. \sysname addresses this gap by capitalizing on the reasoning capability of the \gls{llms} to connect optimization strategies to bottlenecks commonly found in the literature.

Autotuners such as OpenTuner~\cite{ansel2014opentuner} and learning-based tools such as OptFormer~\cite{luo2023optformer} automate code-space exploration via heuristic search or policy learning. However, they rely on manual knob definition or optimize for proxy objectives such as loop unrolling depth or binary size. In contrast, \sysname derives optimization intent directly from multi-source performance diagnostics and maps them to meaningful code transformations using off-the-shelf LLMs. To our knowledge, \sysname is the first framework to integrate diverse analytic sources into an automated code optimization engine.

Andrews et al.’s GPU Kernel Scientist~\cite{AndrewsWitteveen2025} introduces an evolutionary framework that iteratively tests new code versions. This framework implements multiple agents that list all code characteristics, generate experiments to change those characteristics for better performance, and draw on additional knowledge from the internet to refine the experiments. After the experiments are designed, the framework iteratively evaluates them based on runtime and, at the end, provides code that implements optimization. Although this approach is more hands-off and hardware-agnostic, the results are highly constrained. The authors tested the framework on a single application from the AMD Developer Challenge.

KernelBench, introduced by Ouyang et al.~\cite{ouyang2025kernelbench}, provides a framework for evaluating LLMs' ability to generate GPU kernels and optimize them. They train their own fine-tuned model and use an iterative process to achieve speedup. However, their whole experimental setup revolves around generating HPC kernel code. Several studies show that even frontier reasoning models can produce faster, more accurate results. The models actually struggle to make changes and updates to existing code.

%% file: conclusions.tex
This paper introduces \sysname, which transforms today's manual and expert-driven performance optimization task into an automated workflow. To achieve that, \sysname converts multi-source bottleneck diagnostics into structured prompts that guide LLMs to produce evidence-driven code transformations.
Across $\sim$3,410 experiments, \sysname achieves 0.21\%–87.6\% runtime improvement, with fewer than 0.27\% of cases (40/3,410) failing to yield performance improvement. This study demonstrates that LLMs can democratize expert-level GPU performance optimization beyond vendors.

%% file: appendix.tex

\section{Prompt for the \texttt{accuracy} application}
\label{app:accuracy-prompt}

Below is a fully instantiated example of the prompt for the \emph{Accuracy} kernel, precisely reflecting the performance data used in our experiments. The original kernel source code is omitted for brevity.

\vspace{0.5em}

\noindent
\textbf{ROOFLINE SUMMARY:}
\begin{itemize}[leftmargin=*,noitemsep,topsep=0pt]
    \item \textbf{Bound Type}: \textcolor{blue}{memory-bound}
    \item \textbf{Memory Bandwidth}: \textcolor{blue}{30\% of peak}
    \item \textbf{Compute Utilization}: \textcolor{blue}{10\% of peak}
    \item \textbf{Arithmetic Intensity}: \textcolor{blue}{0.5 FLOPs/byte}
    \item \textbf{Profiler Notes}: \textcolor{blue}{Global memory access pattern inefficient; frequent L2 cache misses causing full DRAM fetches.}
\end{itemize}

\noindent
\textbf{STALL ANALYSIS SUMMARY:}
\begin{itemize}[leftmargin=*,noitemsep,topsep=0pt]
    \item \textbf{Memory Dependency Stalls}: \textcolor{blue}{83.2\%, suggesting latency due to inefficient global memory access.}
    \item \textbf{Shared Memory Bank Conflicts}: \textcolor{blue}{10.0\%, indicating non-optimal shared memory usage.}
    \item \textbf{Warp Issue Efficiency}: \textcolor{blue}{0.4 out of 13.6 warps eligible per cycle, implying pipeline congestion or imbalance.}
    \item \textbf{Scheduler Utilization}: \textcolor{blue}{28.6\% actual vs. 100\% theoretical, highlighting execution unit underutilization.}
\end{itemize}

\noindent
\textbf{KEY HARDWARE EVENTS (COUNTERS):}
\begin{itemize}[leftmargin=*,noitemsep,topsep=0pt]
    \item \textbf{lts\_\_t\_sectors\_evict\_first\_lookup\_miss.sum}: \textcolor{blue}{0.189; high L2 cache evictions indicating poor spatial locality or reuse.}
    \item \textbf{smsp\_\_pcsamp\_warps\_issue\_stalled\_mio\_throttle}: \textcolor{blue}{0.209; significant global memory pipeline congestion.}
    \item \textbf{smsp\_\_pcsamp\_warps\_issue\_stalled\_mio\_throttle\_not\_issued}: \textcolor{blue}{0.200; instructions frequently stalled from issue due to memory backpressure.}
\end{itemize}

\noindent
\textbf{TASK INSTRUCTION:}  
Propose optimizations to address the above performance issues. Follow the following instruction:\newline
\texttt{<Instruction 1: Do not rename or duplicate any kernel or function.>\newline
<Instruction 2: More instruction on how to structure the responses, explanations and how to avoid common pitfalls>\newline
..........\newline}
Clearly explain how each suggested optimization relates directly to the identified bottlenecks.

\section{\sysname User Interface}
\begin{figure}[t]
    \centering
    \includegraphics[width=\columnwidth]{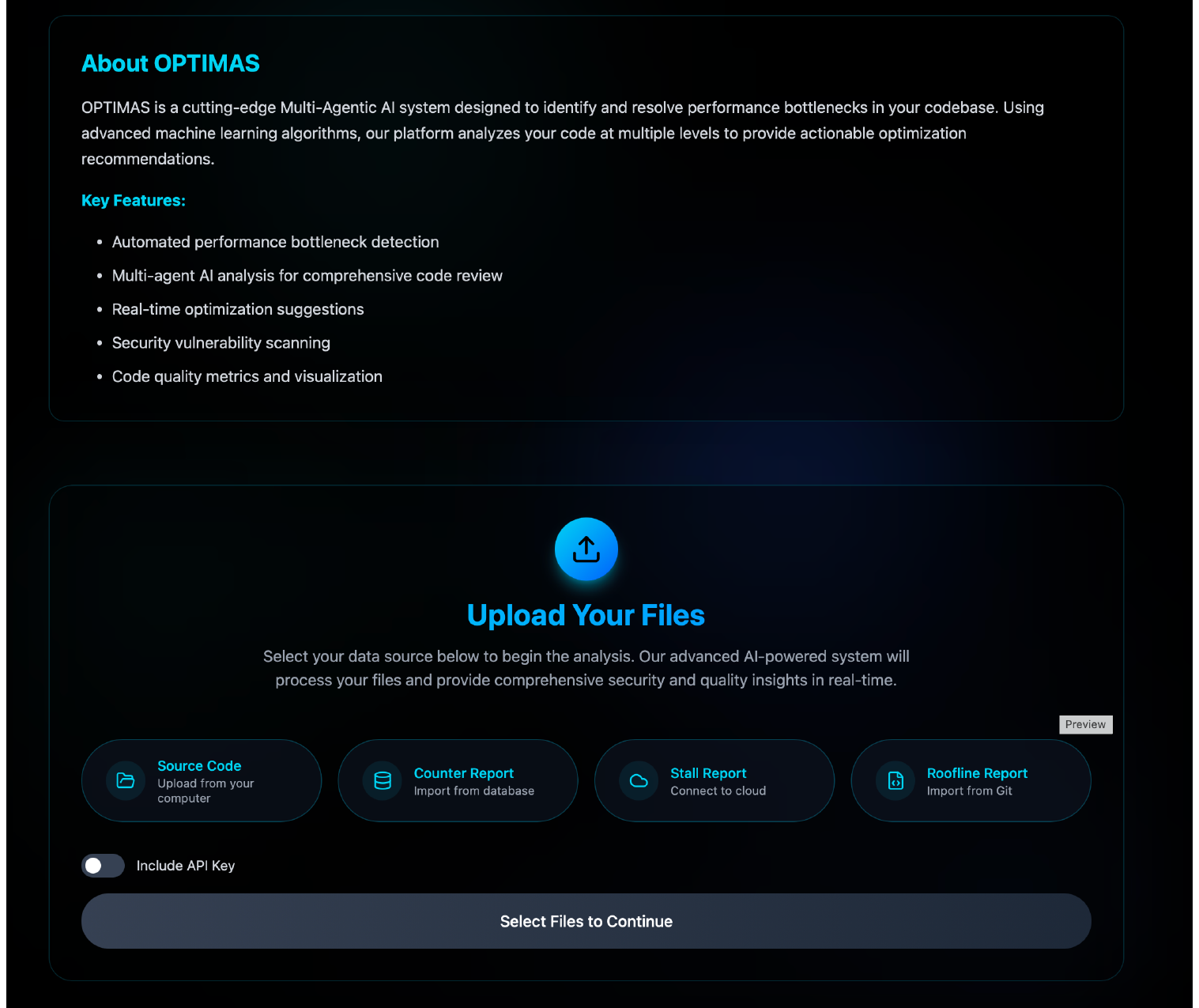}
    \caption{User interface of \sysname.}
    \label{fig:optimas_ui}
\end{figure}
Figure~\ref{fig:optimas_ui} shows the graphical user interface that users interact with to configure and run \sysname. The GUI allows users to upload source code and diagnostics, select analysis options, and inspect optimization results through a unified interface.

\section{Specific Details of the Optimization Recommendations for Each Application}

\input{table_opt_summary}
Table~\ref{tab:default-opt-summary} lists all summary of optimization suggested by each LLM for all of the benchmark applications.

%% file: table_opt_summary.tex
\begin{table*}[t]
\tiny
\caption{
Summarizes the concrete code transformations produced by each LLM under identical performance insights.  
Edits follow standard GPU optimization principles: memory-traffic reduction for streaming kernels (Copy, Add, Triad), stall and reuse mitigation for shared-memory kernels (Shmembench), and coordinated instruction-, memory-, and launch-level tuning for mixed-bottleneck, multi-kernel applications (Accuracy, Sobol, SW4Lite, XSBench).
}
\label{tab:default-opt-summary}
\begin{singlespacing}
\centering
\resizebox{\textwidth}{!}{%
\begin{tabular}{|p{5cm}|p{2cm}|p{5cm}|}
\hline
\textbf{Optimization Summary - GPT5} & \textbf{Optimization Summary - LLama} & \textbf{Optimization Summary - Gemini}\\
\hline

\multicolumn{3}{|l|}{\textbf{Copy, Mul, Add, Triad: 2--3 lines of code.}}\\ \hline
Rewrote kernels with grid-stride loops and UNROLL=4 to reduce per-thread memory traffic, alleviating L1TEX crossbar saturation and L2 tag pressure. Added \_\_launch\_bounds\_\_(TBSIZE) to improve occupancy and latency hiding under high L2 miss rates. &
Applied fused multiply-add (FMA) instructions to reduce floating-point instruction count. &
Rewrote all streaming kernels using grid-stride loops and a fixed number of thread blocks to improve bandwidth utilization and latency hiding. \\
\hline

\multicolumn{3}{|l|}{\textbf{Dot: 19 lines of code.}}\\ \hline
Same as Copy. &
Same as Copy. &
Replaced shared-memory reductions with \_\_shfl\_down\_sync to eliminate divergence.\\
\hline

\multicolumn{3}{|l|}{\textbf{Shmembench: 35 lines of code.}}\\ \hline
Simplified vector assignment (line 28) to reduce instruction count; removed \texttt{\_\_threadfence\_block()} (line 48) to reduce MIO pipeline stalls. &
Privatized shared-memory data into registers and removed unnecessary synchronizations, shifting the kernel from memory-bound to compute-bound. Corrected grid configuration for full data coverage. &
Replaced shared-memory buffers with register privatization (float4 v0–v5), removed \_\_syncthreads(), unrolled loops with \#pragma unroll 32, and fixed launch grid (TOTAL\_BLOCKS/2) to process the full dataset. \\
\hline

\multicolumn{3}{|l|}{\textbf{Accuracy: 41 lines of code.}}\\ \hline
Added \texttt{\_\_restrict\_\_} qualifiers to enable alias-free memory access and reduce uncoalesced loads. &
Reduced synchronization overhead by replacing block-wide barriers with atomicAdd and local registers, targeting Roofline bottlenecks. &
Replaced grid-stride looping with contiguous per-block chunking (rows\_per\_block, start\_row, end\_row) to improve cache locality for Xdata and labelData. Retained CUB BlockReduce while reducing MIO stalls (smsp\_\_pcsamp\_warps\_issue\_stalled\_mio\_throttle). \\
\hline

\multicolumn{3}{|l|}{\textbf{Sobol: 195 lines of code.}}\\ \hline
Precomputed log2(stride) to eliminate repeated \_\_ffs calls, unrolled the first loop to increase ILP, and used \_\_ldg for read-only direction numbers to reduce MIO and short scoreboard stalls. &
Used \_\_int\_as\_float for faster integer--float conversion. &
Replaced regular global stores with inline PTX \texttt{asm("st.global.cg.f32")} to bypass L1 for one-time writes, reducing dram\_\_bytes\_write.sum and l1tex\_\_m\_l1tex2xbar\_write\_bytes.sum traffic. \\
\hline

\multicolumn{3}{|l|}{\textbf{Adjacent: 180 lines of code.}}\\ \hline
Reduced global memory traffic by merging cudaMemcpy calls and using int4 for improved coalescing and cache utilization. &
Reduced L1TEX stalls via memory access pattern optimization. &
Replaced 128-bit vectorized stores with non-temporal 32-bit \_\_stwt stores to bypass L1, reducing L1/L2 contention for streaming outputs. \\
\hline

\multicolumn{3}{|l|}{\textbf{Extrema: 160 lines of code.}}\\ \hline
Improved memory coalescing using int4 and added \_\_syncthreads() to reduce stall\_barrier\_ratio and improve warp scheduling. &
Inserted \_\_syncthreads() in inner loops to ensure correctness and hide latency. &
Introduced shared-memory tiling (extern \_\_shared\_\_ T s\_data[]) for halo caching, dynamic shared memory sizing at launch, and early-exit logic inside comparison loops. \\
\hline

\multicolumn{3}{|l|}{\textbf{addsgd4\_SM: 149 lines of code.}}\\ \hline
Unrolled loops and removed array-offset helper calls (lines 12702--12722) to reduce register pressure and improve locality. &
Removed unused variables and reordered thread indices to reduce memory traffic and improve coalescing. &
Removed unused locals, replaced shared-memory shifts with circular-buffer indexing, inverted the divergent \texttt{for c} loop, and decomposed complex expressions into named intermediates. \\
\hline

\multicolumn{3}{|l|}{\textbf{\swlite: 14{,}017 lines of code.}}\\ \hline
Introduced grid-stride loops to increase eligible warps, applied FMA to replace non-fused FP64 instruction pairs, and added \#pragma unroll. &
Removed unused variables and reordered thread indices to improve coalescing. &
Added \_\_restrict\_\_ to pointer arguments in \texttt{addsgd4\_dev} and \texttt{rhs4center\_dev}, and replaced stencil patterns \texttt{A - 2*B + C} with \texttt{fma(-2.0, B, A) + C}. \\
\hline

\multicolumn{3}{|l|}{\textbf{\xsbench: 81 lines of code.}}\\ \hline
Increased block size to 512 to improve occupancy; added shared memory to reduce global loads; reorganized accesses for coalescing; reduced register usage; used warp-level primitives to minimize divergence; replaced double with single precision; unrolled loops to increase throughput. &
Improved global memory access patterns and reduced uncoalesced loads and stores. &
Restructured the simulation via data sorting, replaced FP instruction sequences with FMA, and removed redundant loads and branches.\\
\hline

\end{tabular}
}
\end{singlespacing}
\end{table*}

%% file: llm-bib.bib
@inproceedings{ansel2014opentuner,
  title={OpenTuner: An extensible framework for program autotuning},
  author={Ansel, Jason and Kamil, Shoaib and Chan, Jonathan Ragan-Kelley Wong and Veeramachaneni, Kalyan and Ziarek, Lukasz and Amarasinghe, Saman and O'Reilly, Martin C},
  booktitle={PACT},
  year={2014},
  pages={303--316},
  doi={10.1145/2628071.2628092}
}

@article{leinhauser2022amdroofline,
  title={Metrics and design of an instruction roofline model for AMD GPUs},
  author={Leinhauser, Matthew and Widera, Ren{\'e} and Bastrakov, Sergei and Debus, Alexander and Bussmann, Michael and Chandrasekaran, Sunita},
  journal={ACM Transactions on Parallel Computing},
  volume={9},
  number={1},
  year={2022}
}

@inproceedings{cummins2025llmcompiler,
  title={{LLM Compiler}: Foundation Language Models for Compiler Optimization},
  author={Cummins, Chris and Seeker, Volker and Grubisic, Dejan and Rozi{\'e}re, Baptiste and Gehring, Jonas and Synnaeve, Gabriel and Leather, Hugh},
  booktitle={CC 2025},
  year={2025},
  pages={141--153}
}

@inproceedings{tramm2014xsbench,
  title={{XSBench}: the development and verification of a performance abstraction for Monte Carlo reactor analysis},
  author={Tramm, John and Siegel, Andrew and Islam, Tanzima and Schulz, Martin},
  booktitle={PHYSOR},
  year={2014}
}


%% file: tzi.bib
@article{deakin2018babelstream,
  author  = {Deakin, Tom and Price, James and Martineau, Matt and McIntosh-Smith, Simon},
  title   = {Evaluating attainable memory bandwidth of parallel programming models via BabelStream},
  journal = {International Journal of Computational Science and Engineering},
  year    = {2018},
  volume  = {17},
  number  = {3},
  pages   = {247--262},
  doi     = {10.1504/IJCSE.2018.095847},
  note    = {Special issue},
}

@article{mess_memory_2024,
  title        = {A Mess of Memory System Benchmarking, Simulation and Application Profiling},
  year         = {2024},
  journal      = {arXiv},
  note         = {arXiv:2405.10170},
}

@inproceedings{naderantahan2021topdown,
  title={Top-Down GPU-Compute Benchmarking using Real-Life Applications},
  author={Naderan-Tahan, M. and others},
  booktitle={IEEE International Symposium on Workload Characterization (IISWC)},
  year={2021}
}

@inproceedings{banday2025role,
  title={On the role of prompt construction in enhancing efficacy and efficiency of llm-based tabular data generation},
  author={Banday, Banooqa and Thopalli, Kowshik and Islam, Tanzima Z and Thiagarajan, Jayaraman J},
  booktitle={ICASSP 2025-2025 IEEE International Conference on Acoustics, Speech and Signal Processing (ICASSP)},
  pages={1--5},
  year={2025},
  organization={IEEE}
}

@misc{Purple,
	author		= {{Lawrence Livermore National Laboratory}},
        title = {The {ASCI} {P}urple {B}enchmark {C}odes},
        note = {\url{http://www.llnl.gov/asci/purple/benchmarks/limited/code_list.html}},
				year={2002}
}

@inproceedings{islam2016a,
  title={A machine learning framework for performance coverage analysis of proxy applications},
  author={Islam, Tanzima Z and Thiagarajan, Jayaraman J and Bhatele, Abhinav and Schulz, Martin and Gamblin, Todd},
  booktitle={SC'16: Proceedings of the International Conference for High Performance Computing, Networking, Storage and Analysis},
  pages={538--549},
  year={2016},
  organization={IEEE}
}

@inproceedings{islam2019toward,
  title={Toward a Programmable Analysis and Visualization Framework for Interactive Performance Analytics},
  author={Islam, Tanzima and Ayala, Alexis and Jensen, Quentin and Ibrahim, Khaled},
  booktitle={2019 IEEE/ACM International Workshop on Programming and Performance Visualization Tools (ProTools)},
  pages={70--77},
  year={2019},
  organization={IEEE}
}

@misc{dashing,
  title={Dashing--A Machine Learning Toolkit for HPC Performance Analysis},
  author={Islam, Tanzima Z. and Ayala, A. and Jensen, Q.},
  howpublished={\url{https://github.com/tzislam/Dashing}}
}

@INPROCEEDINGS{ibrahim2016PCA,
  author={M. F. I. {Ibrahim} and A. A. {Al-Jumaily}},
  booktitle={2016 8th Cairo International Biomedical Engineering Conference (CIBEC)}, 
  title={PCA indexing based feature learning and feature selection}, 
  year={2016},
  volume={},
  number={},
  pages={68-71}}

@inproceedings{chabbi2013effective,
  title={Effective sampling-driven performance tools for GPU-accelerated supercomputers},
  author={Chabbi, Milind and Murthy, Karthik and Fagan, Michael and Mellor-Crummey, John},
  booktitle={Proceedings of the International Conference on High Performance Computing, Networking, Storage and Analysis},
  pages={1--12},
  year={2013}
}

@inproceedings{zaeed2024characterize,
  title={Characterize and Compare the Performance of Deep Learning Optimizers in Recurrent Neural Network Architectures},
  author={Zaeed, Mohammad and Islam, Tanzima Z and Indict, Vladimir},
  booktitle={2024 IEEE 48th Annual Computers, Software, and Applications Conference (COMPSAC)},
  pages={39--44},
  year={2024},
  organization={IEEE}
}

@article{madaan2023learning,
  title={Learning performance-improving code edits},
  author={Madaan, Aman and Shypula, Alexander and Alon, Uri and Hashemi, Milad and Ranganathan, Parthasarathy and Yang, Yiming and Neubig, Graham and Yazdanbakhsh, Amir},
  journal={arXiv preprint arXiv:2302.07867},
  volume={8},
  year={2023},
  publisher={Feb}
}

@article{banerjee2016cmtbone,
author={Tania Banerjee and Jason Hackl and Mrugesh Shringarpure and Tanzima Islam and S Balachandar and Thomas Jackson and Sanjay Ranka},
title={{CMT-Bone — A Proxy Application for Compressible Multiphase Turbulent Flows}},
journal={IEEE 23rd International Conference on High Performance Computing (HiPC)},
year={2016},
number={},
pages={173-182},
doi={10.1109/HiPC.2016.029},
month={Dec},
note={Acceptance rate: 23\%},
url={https://ieeexplore.ieee.org/abstract/document/7839682}}

@inproceedings{wu2021performance,
  title={Performance and energy improvement of ECP proxy app SW4lite under various workloads},
  author={Wu, Xingfu and Taylor, Valerie and Lan, Zhiling},
  booktitle={2021 IEEE/ACM Workshop on Memory Centric High Performance Computing (MCHPC)},
  pages={17--24},
  year={2021},
  organization={IEEE}
}

@misc{lawson2015experimentationprocedureoffloadedminiapps,
      title={Experimentation Procedure for Offloaded Mini-Apps Executed on Cluster Architectures with Xeon Phi Accelerators}, 
      author={Gary Lawson and Vaibhav Sundriyal and Masha Sosonkina and Yuzhong Shen},
      year={2015},
      eprint={1509.02135},
      archivePrefix={arXiv},
      primaryClass={cs.DC},
      url={https://arxiv.org/abs/1509.02135}
}

@misc{cuda_guide2024,
  title        = {CUDA C++ Best Practices Guide},
  author       = {{NVIDIA Corporation}},
  year         = {2024},
  note         = {\url{https://docs.nvidia.com/cuda/cuda-c-best-practices-guide/}}
}

@article{crago2018exposing,
  title={Exposing memory access patterns to improve instruction and memory efficiency in GPUs},
  author={Crago, Neal C and Stephenson, Mark and Keckler, Stephen W},
  journal={ACM Transactions on Architecture and Code Optimization (TACO)},
  volume={15},
  number={4},
  pages={1--23},
  year={2018},
  publisher={ACM New York, NY, USA}
}

@article{Ilic14CRM,
 author = {Ilic, Aleksandar and Pratas, Frederico and Sousa, Leonel},
 title = {Cache-aware Roofline Model: Upgrading the Loft},
 journal = {IEEE Comput. Archit. Lett.},
 issue_date = {January 2014},
 volume = {13},
 number = {1},
 month = jan,
 year = {2014},
 issn = {1556-6056},
 pages = {21--24},
 numpages = {4},
 acmid = {2776841},
 publisher = {IEEE Computer Society},
 address = {Washington, DC, USA},
}

@InProceedings{RooflineScalingTrajectories,
 author = {{Khaled Z. Ibrahim, Samuel Williams, Leonid Oliker}},
 title = {{Roofline Scaling Trajectories: A Method for Parallel Application and Architectural Performance Analysis}},
 booktitle={International Conference on High Performance Computing \& Simulation (HPCS)},
 month = Jun,
 year = 2018,
 }

@article{lee2020roofline,
  title={Roofline-based Data Migration Methodology for Hybrid Memories},
  author={Lee, Jongmin and Lee, Kwangho and Kim, Mucheol and Park, Geunchul and Park, Chan Yeol},
  journal={Journal of Internet Technology},
  volume={21},
  number={3},
  pages={849--859},
  year={2020}
}

@article{williams2009roofline,
  title={Roofline: Aninsightful visual performance model for floating-point programsand multicore architectures},
  author={Williams, S and Waterman, A and Patterson, D},
  journal={Communications of theAssociation for Computing Machinery},
  year={2009}
}

@inproceedings{islam2025data,
  title={Data-Driven Analysis to Understand GPU Hardware Resource Usage of Optimizations},
  author={Islam, Tanzima Z and Marathe, Aniruddha and Schutte, Holland and Zaeed, Mohammad},
  journal={International Journal High Performance Computing Applications (IJHPCA)},
  year={2025},
    note = {Accepted}
}

@inproceedings{meswani2012tools,
  title={Tools for benchmarking, tracing, and simulating SHMEM applications},
  author={Meswani, Mitesh R and Carrington, Laura and Snavely, Allan and Poole, Stephen},
  booktitle={Proceedings Cray user group conference (CUG)},
  year={2012}
}

@techreport{nvidia2023nsight,
  title={Nsight Compute Profiling Guide},
  institution={NVIDIA},
  year={2023}
}

@article{zhou2024star,
  title={Star-Agents: Automatic Data Optimization with LLM Agents for Instruction Tuning},
  author={Zhou, Hang and Tang, Yehui and Qin, Haochen and Yang, Yujie and Jin, Renren and Xiong, Deyi and Han, Kai and Wang, Yunhe},
  journal={Advances in Neural Information Processing Systems},
  volume={37},
  pages={4575--4597},
  year={2024}
}

@misc{chen2021evaluatinglargelanguagemodels,
      title={Evaluating Large Language Models Trained on Code}, 
      author={Mark Chen and Jerry Tworek and Heewoo Jun and Qiming Yuan and Henrique Ponde de Oliveira Pinto and Jared Kaplan and Harri Edwards and Yuri Burda and Nicholas Joseph and Greg Brockman and Alex Ray and Raul Puri and Gretchen Krueger and Michael Petrov and Heidy Khlaaf and Girish Sastry and Pamela Mishkin and Brooke Chan and Scott Gray and Nick Ryder and Mikhail Pavlov and Alethea Power and Lukasz Kaiser and Mohammad Bavarian and Clemens Winter and Philippe Tillet and Felipe Petroski Such and Dave Cummings and Matthias Plappert and Fotios Chantzis and Elizabeth Barnes and Ariel Herbert-Voss and William Hebgen Guss and Alex Nichol and Alex Paino and Nikolas Tezak and Jie Tang and Igor Babuschkin and Suchir Balaji and Shantanu Jain and William Saunders and Christopher Hesse and Andrew N. Carr and Jan Leike and Josh Achiam and Vedant Misra and Evan Morikawa and Alec Radford and Matthew Knight and Miles Brundage and Mira Murati and Katie Mayer and Peter Welinder and Bob McGrew and Dario Amodei and Sam McCandlish and Ilya Sutskever and Wojciech Zaremba},
      year={2021},
      eprint={2107.03374},
      archivePrefix={arXiv},
      primaryClass={cs.LG},
      url={https://arxiv.org/abs/2107.03374}
}

@INPROCEEDINGS{jin2023a,
  author={Jin, Zheming and Vetter, Jeffrey S.},
  booktitle={2023 IEEE International Symposium on Performance Analysis of Systems and Software (ISPASS)}, 
  title={A Benchmark Suite for Improving Performance Portability of the SYCL Programming Model}, 
  year={2023},
  volume={},
  number={},
  pages={325-327},
  doi={10.1109/ISPASS57527.2023.00041}}

@article{zhou2021measurement,
  title={Measurement and analysis of GPU-accelerated applications with HPCToolkit},
  author={Zhou, Keren and Adhianto, Laksono and Anderson, Jonathon and Cherian, Aaron and Grubisic, Dejan and Krentel, Mark and Liu, Yumeng and Meng, Xiaozhu and Mellor-Crummey, John},
  journal={Parallel Computing},
  volume={108},
  pages={102837},
  year={2021},
  publisher={Elsevier}
}

@inproceedings{uhsadel2008exploiting,
  title={Exploiting hardware performance counters},
  author={Uhsadel, Leif and Georges, Andy and Verbauwhede, Ingrid},
  booktitle={2008 5th Workshop on Fault Diagnosis and Tolerance in Cryptography},
  pages={59--67},
  year={2008},
  organization={IEEE}
}

@inproceedings{zhou2020tools,
  title={Tools for top-down performance analysis of GPU-accelerated applications},
  author={Zhou, Keren and Krentel, Mark W and Mellor-Crummey, John},
  booktitle={Proceedings of the 34th ACM International Conference on Supercomputing},
  pages={1--12},
  year={2020}
}

@inproceedings{zhou2021gpa,
  title={Gpa: A gpu performance advisor based on instruction sampling},
  author={Zhou, Keren and Meng, Xiaozhu and Sai, Ryuichi and Mellor-Crummey, John},
  booktitle={2021 IEEE/ACM International Symposium on Code Generation and Optimization (CGO)},
  pages={115--125},
  year={2021},
  organization={IEEE}
}

@misc{amd_roofline_micro2022,
  author       = {{Advanced Micro Devices, Inc.}},
  title        = {MI200 Performance Counters and Metrics},
  year         = {2024},
  howpublished = {\url{https://rocm.docs.amd.com/en/docs-6.0.0/conceptual/gpu-arch/mi200-performance-counters.html}},
  note         = {Accessed: 2025-04-14}
}

@article{li2023starcoder,
  title={StarCoder: May the source be with you!},
  author={Li, Raymond and Ben Allal, Loubna and Zi, Yangtian and Muennighoff, Niklas and Kocetkov, Denis and Mou, Chenghao and Marone, Marc and Akiki, Christopher and Li, Jia and Chim, Jenny and others},
  journal={arXiv preprint arXiv:2305.06161},
  year={2023},
  url={https://arxiv.org/abs/2305.06161}
}

@article{roziere2023code,
  title={Code Llama: Open Foundation Models for Code},
  author={Rozi{\`e}re, Baptiste and Gehring, Jonas and Gloeckle, Fabian and Sootla, Sten and Gat, Itai and Tan, Xiaoqing Ellen and Adi, Yossi and Liu, Jingyu and Sauvestre, Romain and Remez, Tal and others},
  journal={arXiv preprint arXiv:2308.12950},
  year={2023},
  url={https://arxiv.org/abs/2308.12950}
}

@article{nichols2024rlpf,
  title={Performance-Aligned LLMs for Generating Fast Code},
  author={Nichols, Daniel and Polasam, Pranav and Menon, Harshitha and Marathe, Aniruddha and Gamblin, Todd and Bhatele, Abhinav},
  year={2024},
  howpublished={arXiv preprint},
  url={https://pssg.cs.umd.edu/assets/papers/2024-04-rlpf-arxiv.pdf}
}

@article{nichols2024hpccoder,
  title={{HPC-Coder}: Modeling Parallel Programs using Large Language Models},
  author={Nichols, Daniel and Marathe, Aniruddha and Menon, Harshitha and Gamblin, Todd and Bhatele, Abhinav},
  year={2024},
  howpublished={arXiv/ISC preprint},
  url={https://pssg.cs.umd.edu/assets/papers/2024-05-hpc-llm-isc.pdf}
}

@article{luo2023optformer,
  title={OptFormer: Schedule Optimization as Program Generation},
  author={Luo, Yukuo and Zheng, Yanqi and Deng, Jiacheng and Zhang, Tianjun and Su, Hang and Ermon, Stefano and Ye, Yuke and Zhao, Tianshi and Wu, Yuxin},
  journal={arXiv preprint arXiv:2305.10765},
  year={2023},
  url={https://arxiv.org/abs/2305.10765}
}

@INPROCEEDINGS{islam2019dashing,
  author={Islam, Tanzima and Ayala, Alexis and Jensen, Quentin and Ibrahim, Khaled},
  booktitle={2019 IEEE/ACM International Workshop on Programming and Performance Visualization Tools (ProTools)}, 
  title={Toward a Programmable Analysis and Visualization Framework for Interactive Performance Analytics}, 
  year={2019},
  volume={},
  number={},
  pages={70-77},
  doi={10.1109/ProTools49597.2019.00015}
}

@inproceedings{du2025turn,
  title     = {Optimizing Temperature for Language Models with Multi-Sample Inference},
  author    = {Weihua Du and Yiming Yang and Sean Welleck},
  booktitle = {Proceedings of the International Conference on Machine Learning (ICML)},
  year      = {2025}
}

@misc{AndrewsWitteveen2025,
  title        = {GPU Kernel Scientist: An LLM-Driven Framework for Iterative Kernel Optimization},
  author       = {Andrews, Martin and Witteveen, Sam},
  year         = {2025},
  howpublished = {arXiv preprint arXiv:2506.20807v2},
  note         = {\url{https://arxiv.org/abs/2506.20807v2}}
}

@misc{nvidia_cccl_adjacent_difference,
  title        = {cub::DeviceAdjacentDifference — CUDA Core Compute Libraries},
  howpublished = {\url{https://nvidia.github.io/cccl/cub/api/structcub_1_1DeviceAdjacentDifference.html}},
  note         = {Accessed: \today},
}

@misc{Thompson2021_cuSignal,
  author       = {Thompson, Adam},
  title        = {Accelerated Signal Processing with cuSignal},
  howpublished = {NVIDIA Technical Blog},
  month        = mar,
  year         = {2021},
  note         = {\url{https://developer.nvidia.com/blog/accelerated-signal-processing-with-cusignal/} (accessed: \today)},
}

@article{ouyang2025kernelbench,
  title        = {KernelBench: Can LLMs Write Efficient GPU Kernels?},
  author       = {Ouyang, Anne and Guo, Simon and Arora, Simran and Zhang, Alex L. and Hu, William and Re{\'e}, Christopher and Mirhoseini, Azalia},
  journal      = {arXiv preprint arXiv:2502.10517},
  year         = {2025},
  url          = {https://arxiv.org/abs/2502.10517}
}
